# COMMISSIONING AND OPERATION OF FAST ELECTRON LINAC AT FERMILAB*

A. Romanov†, C. Baffes, D. R. Broemmelsiek, K. Carlson, D. J. Crawford, N. Eddy, D. Edstrom, Jr, E. R. Harms, J. Hurd, M. Kucera, J. Leibfritz, I. Rakhno, J. Reid, J. Ruan, J. Santucci, V. Shiltsev, G. Stancari, R. Thurman-Keup, A. Valishev, A. Warner, Fermilab, [60510] Batavia, USA

*Abstract*

We report results of the beam commissioning and first operation of the 1.3 GHz superconducting RF electron linear accelerator at Fermilab Accelerator Science and Technology (FAST) facility. Construction of the linac was completed and the machine was commissioned with beam in 2017. The maximum total beam energy of about 300 MeV was achieved with the record energy gain of 250 MeV in the ILC-type SRF cryomodule. The photoinjector was tuned to produce trains of 200 pC bunches with a frequency of 3 MHz at a repetition rate of 1 Hz. This report describes the aspects of machine commissioning such as tuning of the SRF cryomodule and beam optics optimization. We also present highlights of an experimental program carried out parasitically during the two-month run, including studies of wake-fields, and advanced beam phase space manipulation.

## INTRODUCTION

Construction of a 300 MeV electron linac based on superconducting RF (SRF) technology has been completed at the FAST facility [1, 2]. The primary use of the machine will be to provide 150 MeV electrons for injection into the Integrable Optics Test Accelerator (IOTA) ring currently being built to test novel techniques of non-linear integrable optics [3] and space-charge compensation [4] as well as other advanced accelerator technologies. The low duty factor of the electron injector required for IOTA (i.e. a single bunch to be injected on demand every few minutes) allows for a parallel beamline-based test program to make use of the large, stable operational parameter space summarized in Table 1. The injector itself consists of the following major components: a photoinjector-based electron gun [5], a 25-meter-long low-energy (≤50 MeV) pre-accelerator [6], an 8-cavity ILC-type cryomodule [7], and a 100-meter-long high-energy (≤300 MeV) beamline – see Fig. 1 from [1]. Construction of the IOTA ring at the end of the high energy section is in its final stages. Here we report results of the commissioning of first beam to the maximum energy of 300 MeV and of matched beam at 150 MeV to the high energy beam absorber (HEA).

## THE FAST BEAMLINE

The beamline starts with a photocathode in a normal-conducting electron RF gun with 1.5-cell copper cavity operating at 1.3 GHz. A train of UV pulses from the drive laser strikes a $Cs_2Te$-coated Mo cathode in the gun cavity resulting in a train of electron bunches with the bunch charge of up to 3.2 nC. The cathode and the gun cavity are surrounded by two solenoids that create a focusing field to counteract space charge forces. The nominal setup assumes zero magnetic field on the cathode, but individual control over the solenoids allows operation with magnetized beams.

Beam from the gun passes through a short (~1 m) low-energy diagnostic section before acceleration in two consecutive superconducting RF structures, referred to as the capture cavities, CC1 and CC2. Each capture cavity in its own cryostat is a 9-cell, 1.3 GHz (L-Band), Nb accelerating structure nominally cooled to 2 K. Following acceleration, the electron beam passes through the low-energy beam transport section, which includes steering and focusing elements, an optional chicane for bunch compression and beam transforms, as well as a set of diagnostic devices. It is then either directed into the low-energy absorber or to the high-energy section. The high-energy section begins with a TESLA Type IV ILC-style eight cavity 1.3 GHz SRF cryomodule, referred to as CM2, has been conditioned to an average accelerating gradient of about 31.5 MeV/m [8, 9]. This provided a 255 MeV energy gain to the beam before it was sent to the HEA.

Table 1: Summary of FAST Beam Parameters

| Parameter | FAST Value |
|---|---|
| Beam Energy to LE Absorber | 20 – 50 MeV |
| Beam Energy to HE Absorber | 100 – 300 MeV |
| Bunch charge | <10 fC – 3.2 nC |
| Bunch Frequency | 0.5 – 9 MHz |
| Macropulse frequency | 1 – 5 Hz |
| Macropulse duration | < 1 ms |
| Normalized emittance (0.1nC/bunch) | 0.6 um |
| Number of trims (X & Y) | 23 & 23 |
| Number of BPMs | 26 |
| Quads (LE & HE) | 14 & 22 |
| Beam imaging stations | 17 |
| Crosses for insertion devices | 20 |

## HIGH ENERGY RUN OVERVIEW

*Cryomodule tuning*

The optimization of RF phases between the gun, CC1 and CC2 is fairly simple, given that each has its own low-

___
* Work supported by the DOE contract No.DEAC02-07CH11359 to the Fermi Research Alliance LLC.
† aromanov@fnal.gov

This manuscript has been authored by Fermi Research Alliance, LLC under Contract No. DE-AC02-07CH11359 with the U.S. Department of Energy, Office of Science, Office of High Energy Physics.

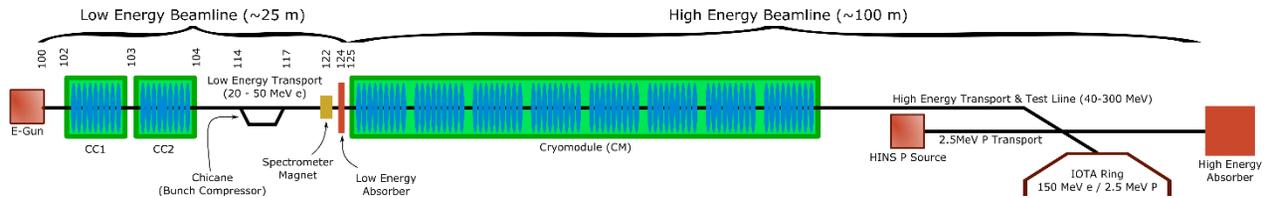

Figure 1: The FAST electron linac: low energy beamline was commissioned to the beam energies up to 52.5 MeV [6] and the high-energy transport section is now operational up to 300 MeV.

level RF (LLRF) and high-level RF (HLRF) systems. Therefore, maximizing the total energy gain is relatively easy with only three cavities. This is not the case for CM2, in which 8 cavities share a single LLRF and HLRF system, resulting in a vector sum that can rapidly become difficult to deconvolute as each relative phase will impact the remaining cavities.

To find the optimal position for the mechanical phase tuners, available for each cavity in the CM2, the FAST operations group relied on a beam-based tuning method. As seen in the example for CM2 cavity no. 1, shown in Fig. 2, the cavity gradient was monitored with a digitizer. The relative timing to the low energy section resulted in the bunch train entering CM2 near digitizer bin 765, as seen on the traces shown in the example. Beam loading could be easily seen on the "flat-top", which in the absence of feedback has a slight negative slope. Adjustment of either the LLRF phase or the phase delay tuning motor could be used to shift the RF phase change corresponding to the beam loading. A phase shift over the complete RF cycle includes both maximum deceleration and maximum acceleration through the cavity (Fig. 3), the latter corresponding to the maximum beam loading. In order to maximize the acceleration for all cavities to a LLRF phase set-point of 0, the phase delay tuning motor was adjusted to maximize the peak beam loading in the cavity.

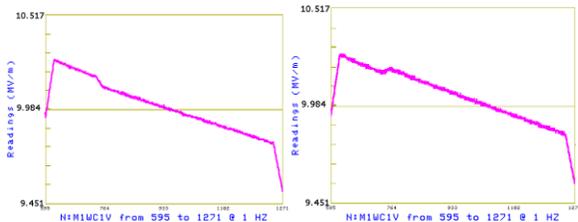

Figure 2: RF gradient readbacks of cryomodule cavity #1 with feedback off. Plots show RF flattops with beam loadings for in-phase (left) and out of phase trains.

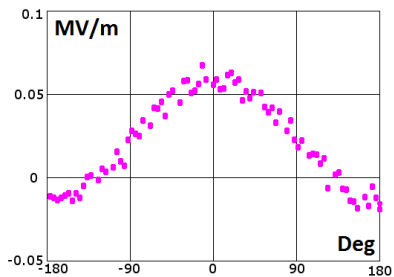

Figure 3: Beam loading effect dependence on the RF phase.

This process was repeated for the remaining seven cavities. To find the ideal position for each of the tuners, the gradient with beam loading was subtracted from a point on the gradient curve before beam injection, and the maximum position was placed roughly at 0° for the LLRF phase setting. A more rigorous tune could be performed by taking the change of gradient slope into account through this tune, but the gradient applied to the cavity, closed loop LLRF operation, and Lorentz Force Detuning compensation will also contribute to any remaining offset.

Once all eight CM2 phase delay tuners were aligned, the beam was accelerated down the high energy beamline to the energy of 298±5 MeV as measured after the 15-degree dipole, which was used as the high energy spectrometer. There are two sources of error in the energy measurement. First is the uncertainty in the beam angles before and after the dipole, which are estimated to be about 2 mrad for the total bend angle of 15 degrees. The second is the uncertainty of magnetic field calibration of the magnet at the level of 1%. In the maximum energy operation, the beam after the low-energy section had the energy of 42.9±0.2 MeV, while the cryomodule accelerated by an additional 255 MeV.

### Lattice tuning

From the practical point of view, FAST tuning was split in two logical steps. First, the front-end of the beamline with the capture cavities was tuned to achieve minimum beam shape distortions at the point where the beam energy becomes high enough to suppress space-charge induced degradation of emittances. At the second step, the available quadrupoles and trim magnets were used to match the beam parameters to the specific requirements of the ongoing experiment.

Three methods were used at FAST to measure the beam parameters after the second capture cavity. The quickest method is to use two sets of insertable slits with an online image analysis tool (Fig. 4) [10]. Second, a quadrupole scan was used (Fig. 5). Table 2 shows the comparison of beam parameters derived from the two mentioned methods for the same beam with a bunch charge of 100 pC. Finally, beam phase space tomography can be used for the detailed analysis of the particle distribution [11].

One of the goals of the commissioning was to tune the beamline lattice for 150 MeV beams to the HEA. It was done to ensure close to lossless transmission of the beam from the gun to the HEA and to make sure that FAST can deliver bunches suitable for the injection into IOTA (Fig. 6 and Fig. 7).

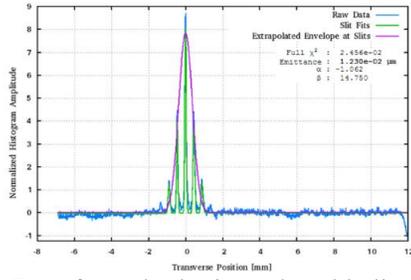

Figure 4: Data from the horizontal multi-slits emittance measurements.

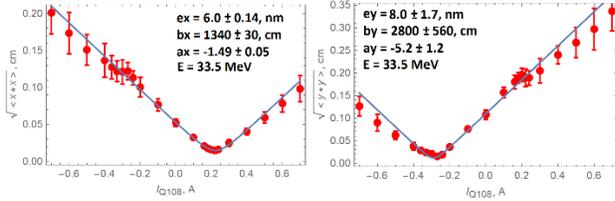

Figure 5: Quadrupole scan data (red circles) and fits (blue lines) for horizontal size (left) and vertical size (right).

Table 2: Comparison of Beam Parameters Measured With Multi-Slits and a Quad Scan

| Parameter | Quad scan | Multi-slits |
|---|---|---|
| $\varepsilon_y$, geometric | 8.0±1.7, nm | 12±2, nm |
| $\beta_y$ | 2800±560, cm | 1475±150, cm |
| $\alpha_y$ | -5.2±1.2 | -1.06±0.2 |

After the initial manual tuning of propagation of the beam to the HEA, alternating steps of beam-based lattice and trajectory corrections were made. The trajectory correction was aimed at minimizing beam offsets from the magnetic axis of quadrupoles. The analysis and correction of the focusing optics was done using the LOCO method [12]. An important aspect of the lattice tuning was related to the proper correction of the dispersion achromat in the dogleg formed by the two 15-degree dipoles (at around 110-120m in Fig. 7). The variation of gradient in CC2 was used to produce the beam energy change. Small but non-zero trajectory errors in CC2 created a trajectory wave, which was accounted for based on the data from BPMs before the achromat. The resulting beam envelopes are presented in Fig. 8 and Fig. 9.

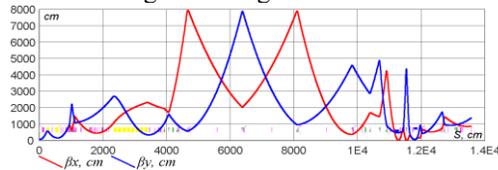

Figure 6: Beta-functions for the FAST linac at 150 MeV

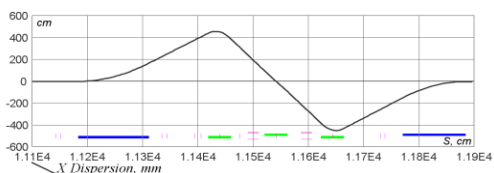

Figure 7: Dispersion achromat in the dogleg of the FAST linac.

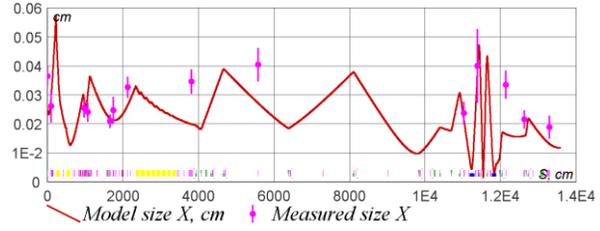

Figure 8: Data Model beam envelope (red line) and measured beam sizes (magenta dots) for horizontal plane.

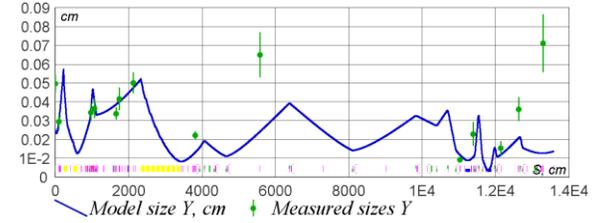

Figure 9: Data Model beam envelope (blue line) and measured beam sizes (green dots) for vertical plane.

*Experimental program*

Operations with two eight-hour shifts per day allowed for combined active cryomodule and beamline commissioning along with several experiments.

Studies of high order modes (HOM) were done to ensure that numerical simulations give reliable predictions for such complicated RF components as the Tesla-style cavities [13]. HOM studies also provided additional information about beam alignment relative to the electromagnetic axis of capture cavities.

Synchrotron radiation from the 15-degree dipole was used for the testing apparatus for the precision experiments at IOTA [14, 15]. Studies of the THz radiation from longitudinally shaped bunches were done using the chicane and ceramic break in the low energy part of FAST [16]. Another study was dedicated to the problem of the front-end tuning in the wide range of required parameters using machine learning algorithms [17].

Finally, two experiments targeted fully coupled beam dynamics. One is the 4D beam phase space tomography [11]. The other study was aimed at the understanding of round-to-flat beam transformations of beams magnetised at the cathode using 3 skew-quadrupoles located after the second capture cavity [18].

## CONCLUSION

We have successfully commissioned the high energy section of the FAST linac and met two major goals: 1) it was demonstrated that the linac can produce 150 MeV beams suitable for all IOTA experiments with electrons, and 2) the ILC-type 8-cavity 1.3 GHz SRF cryomodule has reached the energy gain of 255±5 MeV, such that the FAST beam attained the maximum total energy of 298 MeV. In addition, several beam experiments were successfully done in collaboration with external and internal research groups and in parallel with the fulfilment of the major goals of the facility 2017 run.